\documentclass{jnmp}

\usepackage{amsmath}

\begin{document}

\renewcommand{\evenhead}{A~Karasu(Kalkanl{\i}) and H~Y{\i}ld{\i}r{\i}m}
\renewcommand{\oddhead}{On the Lie Symmetries of Kepler--Ermakov Systems}

\thispagestyle{empty}

\FirstPageHead{9}{4}{2002}{\pageref{karasu-firstpage}--\pageref{karasu-lastpage}}{Article}

\copyrightnote{2002}{A~Karasu(Kalkanl{\i}) and H~Y{\i}ld{\i}r{\i}m}

\Name{On the Lie Symmetries\\ of Kepler--Ermakov Systems}
\label{karasu-firstpage}

\Author{Ay\c{s}e KARASU(KALKANLI) and Hasan YILDIRIM}

\Address{Department of Physics, Middle East Technical University, 06531 Ankara, Turkey\\
E-mail: akarasu@metu.edu.tr}

\Date{Received February 12, 2002; Revised May 4, 2002; Accepted
May, 2002}

\begin{abstract}
\noindent
In this work, we study the Lie-point symmetries of Kepler--Ermakov
systems presented by C~Athorne in {\it J.~Phys.}~{\bf A24}
(1991), L1385--L1389.
We determine the forms of arbitrary
function $H(x,y)$ in order to find the members of this class possessing
the ${\bf sl}(2,\mathbb{R})$ symmetry and a Lagrangian. We show that these
systems are usual Ermakov systems with the frequency function depending
on the dynamical variables.
\end{abstract}

\section{Introduction}

The basic idea behind the symmetry methods is to reduce the order of
differential equations under considerations as much as possible so
that the integration can be done easily~[11]. It is well known that
a dynamical system is integrable if a sufficient number of
sufficiently simple (e.g.~polynomial) independent first integrals
can be found. The knowledge of a first integral
amounts to the reduction of the order of the integration procedure
by one. For systems of~$N$ second order differential equations this
order has to be counted as~$2N$.
Ermakov systems are time-dependent dynamical
systems~[9]. They contain one arbitrary function of time, the so-called frequency function,
and two arbitrary homogeneous functions of dynamical variables.
A trend in the latest developments on the
subject is to focus attention on some special features of subclasses
of generalized Ermakov systems in which the frequency function may
depend on time, the dynamical variables and their derivatives~[5, 6, 8].
These subclasses may be tailored to suit some
particular applications or special purposes. A central feature of
Ermakov systems is their property of always having a first integral~[4].
This invariant plays a central role in the linearization of Ermakov
systems~[2, 6, 10].
A class of dynamical systems was presented by Athorne~[1] which
can be regarded as perturbations of the
classical Kepler problem or of an autonomous Ermakov system.
Kepler--Ermakov systems generalize the usual Ermakov systems
while preserving the property of being amenable to linearization~[1, 3, 6].
In the next section, we study the Lie point symmetries of Kepler--Ermakov
systems. We find the Kepler--Ermakov--Lagrangian systems possessing the
${\bf sl}(2,\mathbb{R})$ symmetry.

\section{Symmetry generators}

Systems of second order differential
equations often appear in classical
mechanics. So we use the notations
common in that field~[11]. We consider the generalized
coordinates $q^{a}$ as dependent variables, $t$ as the
independent variable, and denote $dq^{a}/dt$ by $\dot{q}^{a}$. Then,
\begin{equation}
\ddot{q}^{a} = w^{a} \left(q^{i}, \dot{q}^{i}, t\right), \qquad
a,i = 1, \ldots, N,
\end{equation}
is the general form for a system
of second order differential equations. The corresponding linear partial
differential equation of first order is
\begin{equation}
{\bf A} f = \left(\frac{\partial}{\partial t} + \dot{q}^{a}
\frac{\partial}{\partial q^{a}} + w^{a}
\left(q^{i}, \dot{q}^{i}, t\right)
\frac{\partial}{\partial \dot{q}^{a}}\right) f = 0,
\end{equation}
which admits $2N$ functionally independent solutions
$\varphi^\alpha = \varphi^\alpha \left(q^{a}, \dot{q}^{a}, t\right)$
that are
first integrals of the system~(1). So every solution of~(1) can locally be written as
$q^{a} = q^{a}(\varphi^\alpha , t)$, where $d\varphi^\alpha/dt=0$.

When the operator ${\bf A}$ is written in the form
of~(2), we can write
the infinitesimal generator of the Lie point symmetry
admitted by~(1) as
\begin{equation}
{\bf X} = \xi \left(q^{i}, t\right) \frac{\partial}{\partial t} +
\eta^{a} \left(q^{i}, t\right) \frac{\partial}{\partial q^{a}}.
\end{equation}
Its extension is
\begin{equation}
\dot{{\bf X}} = \xi \frac{\partial}{\partial t} + \eta^{a}
\frac{\partial}{\partial q^{a}} + \left(\frac{d \eta^{a}}{dt} - \dot{q}^{a}
\frac{d \xi}{dt}\right)\frac{\partial}{\partial \dot{q}^{a}},
\end{equation}
and the symmetry condition is
\begin{equation}
[\dot{{\bf X}}, {\bf A}] = \lambda {\bf A},
\end{equation}
where $\lambda$ is a non-constant factor, in general.
By writing~(5) explicitly and matching the terms
on both sides, we obtain the following equations
\begin{gather}
- {\bf A} \xi=- \frac{d \xi}{dt} = \lambda, \\
\dot{{\bf X}} w^{a}={\bf A}\left(\frac{d \eta^{a}}{dt} - \dot{q}^{a}
\frac{d \xi}{dt}\right)- w^{a}\frac{d \xi}{dt}.
\end{gather}
If (7) is written in full, it is
\begin{gather}
 \xi w^{a}_{t} + \eta^b w^{a}_{b} +
(\eta_{t}^b + \dot{q}^c \eta_{c}^b -
\dot{q}^b \xi_{t} - \dot{q}^b \dot{q}^c \xi_{c})
\frac{\partial w^{a}}{\partial \dot{q}^b}
 + 2 w^{a} (\xi_{t} + \dot{q}^b \xi_{b}) + w^b (\dot{q}^{a}
\xi_{b} -\eta^{a}_{b}) \nonumber\\
\qquad {}+ \dot{q}^{a} \dot{q}^b \dot{q}^c \xi_{bc}
 + 2 \dot{q}^{a} \dot{q}^c \xi_{tc} - \dot{q}^c \dot{q}^b
\eta^{a}_{bc} + \dot{q}^{a} \xi_{tt} - 2 \dot{q}^b \eta^{a}_{tb} -
\eta^{a}_{tt} = 0,
\end{gather}
where the  subscripts $b,c=1, \ldots,N$, and $t$ denote partial
differentiations with respect to generalized coordinates and time.

Kepler--Ermakov systems are defined [1] as the system of equations
\begin{gather}
\ddot{x} + w^2 (t) x  =  - \frac{x}{r^3} H + \frac{1}{x^3}
f (y/x),\nonumber\\
 \ddot{y} + w^2 (t) y  =   - \frac{y}{r^3} H + \frac{1}{y^3}
g (y/x),
\end{gather}
where $H$ is a function of $x$, $y$ and $r = \left(x^2 + y^2\right)^{1/2}$.
Here $f$ and $g$ are arbitrary functions of the
indicated arguments and $w$ is an
arbitrary function of time. If $H$, $f$ and $g$ are zero,
we have a time-dependent harmonic oscillator
which occurs in many practical contexts.
In the case that $H$ is taken to be zero
we have generalized Ermakov systems.

In order to calculate the Lie point symmetries of
system~(9) we consider the generator
of the group of point transformations
\begin{equation}
{\bf X} = \xi (x,y,t) \frac{\partial}{\partial t} + \eta_1
(x,y,t) \frac{\partial}{\partial x}
+ \eta_2 (x,y,t) \frac{\partial}{\partial y}.
\end{equation}
By inserting (9) into the symmetry condition
(8) we obtain a polynomial equation
in $\dot{x}$
and $\dot{y}$. The coefficients of all monomials
of the form $\dot{x}^m \dot{y}^n$ must be identically zero. This
yields the following system of partial differential equations
satisfied by $\xi$, $\eta_1$ and $\eta_2$,
\begin{gather}
 \xi_{xx}  = \xi_{yy} = \xi_{xy} = 0,\qquad
\eta_{1 xx} - 2 \xi_{tx} = 0,\qquad  \eta_{1xy} - \xi_{ty} = 0,\nonumber\\
\eta_{2xy} - \xi_{tx} = 0,\qquad  \eta_{2 yy} - 2 \xi_{ty} = 0,\qquad
\eta_{1 yy} = \eta_{2 xx} = 0,
\end{gather}
where we use subscripts to denote partial
derivatives.
The solutions of equations (11) can be written as
\begin{gather}
\xi  =  \kappa (t) x + \delta (t) y + \sigma (t), \qquad
\eta_1  =  \alpha_1 (x,t) y +\beta_1 (x,t),\nonumber\\
\eta_2  =  \alpha_2 (y,t) x + \beta_2 (y,t),
\end{gather}
where the functions $\kappa$, $\delta$, $\sigma$, $\alpha_1$, $\beta_1$, $\alpha_2$
and $\beta_2$ must satisfy the conditions
\begin{gather}
 \alpha_{1xx} = \alpha_{2 yy} = 0,\qquad
 \beta_{1xx} - 2 \dot{\kappa} = 0,\qquad
\beta_{2yy} - 2 \dot{\delta} =0,\nonumber\\
 \alpha_{1x} - \dot{\delta} = 0, \qquad
\alpha_{2y} - \dot{\kappa} = 0.
\end{gather}
The solutions of these equations are
\begin{gather}
\alpha_1 (x,t)  =  \dot{\delta} x + \phi_1 (t), \qquad
\alpha_2 (y,t)  =  \dot{\kappa} y + \phi_2 (t),\nonumber\\
\beta_1 (x,t)  =  \dot{\kappa} x^2 + \phi_3 (t) x + \phi_5 (t), \qquad
\beta_2 (y,t)  =  \dot{\delta} y^2 + \phi_4 (t) y + \phi_6 (t).
\end{gather}
Substituting (12) and (14) into the symmetry condition (8)
and setting the coefficients of
the terms linear in $\dot{x}$ and $\dot{y}$ to
zero we obtain
\begin{gather}
\phi_3 (t) =  ( \dot{\sigma} - c_1)/2, \qquad
\phi_4 (t)  =  (\dot{\sigma} - c_2)/2, \qquad
\kappa  =  \delta = 0.
\end{gather}
We note that, up to here,
$f$, $g$, $H$ and $w$ are arbitrary functions of their arguments.
Inserting (15) into the rest of (8), we
find that
\begin{gather}
c_1 = c_2 = 0,\qquad
\phi_1 = \phi_2 = \phi_5 = \phi_6 = 0,\\
(x H_x + y H_y + 2H) \frac{\dot{\sigma}}{r^3} +
\stackrel{\ldots}{\sigma} + 4 \dot{w} w \sigma +
4 \dot{\sigma} w^2 = 0.
\end{gather}
The last equation implies that the
function $H(x,y)$ is not completely arbitrary but must be a
function on the integral surface
of the partial differential equation
\begin{equation}
x H_x  + y H_y + 2H = C \left(x^2 + y^2\right)^{3/2}
\end{equation}
where $C$ is a constant. As a special case, for example, if $C$
is zero, $H$ must be in the form
\begin{equation}
H =-\frac{h(x/y)}{y^2},
\end{equation}
where $h$ is an arbitrary function of its argument
and the function $\sigma (t)$ must be the solution
of
\begin{equation}
\stackrel{\ldots}{\sigma} + 4 \dot{w} w \sigma  + 4 \dot{\sigma} w^2 = 0
\end{equation}
which is the third order form of Ermakov--Pinney equation.
Then equations (9) have the
symmetry
\begin{equation}
{\bf X} = \sigma (t) \frac{\partial}{\partial t} + \frac{1}{2}
\dot{\sigma} (t) x \frac{\partial}{\partial x} + \frac{1}{2}
\dot{\sigma} (t) y \frac{\partial}{\partial y}
\end{equation}
which is also a symmetry of
\begin{equation}
x \ddot{y} - y \ddot{x} - \frac{x}{y^3} g(y/x) +
\frac{y}{x^3} f(y/x) = 0
\end{equation}
obtained [4, 8] by eliminating $w(t)$ and $H$ terms in~(9).
If $w \neq 0$, equation (20) can be integrated once to
obtain, $\sigma \ddot{\sigma}-\frac{1}{2}\dot{\sigma}^2+
2\sigma^2 w^2=c_1$ which can be reduced to the Pinney
equation, $\ddot{\rho}+w^2\rho = \frac{c_2}{\rho^3}$,
by the change of variable $\sigma=\rho^2(t)$.
On the other hand,
equa\-tion~(20) can be solved easily if $w =0$.
Hence the system (9) with $w = 0$ has
the symmetry
\begin{equation}
{\bf X}  =  \left(c_1 t^2 + c_2 t + c_3\right) \frac{\partial}{\partial t} +
\left(c_1 t x + \frac{c_2}{2} x\right) \frac{\partial}{\partial x}
 + \left(c_1 t y + \frac{c_2}{2} y\right) \frac{\partial}{\partial y}
\end{equation}
which splits into three
components
\begin{gather}
G_1  =  t^2 \frac{\partial}{\partial t} + t x
\frac{\partial}{\partial x} + ty
\frac{\partial}{\partial y},\qquad
G_2  =  t \frac{\partial}{\partial t} + \frac{x}{2}
\frac{\partial}{\partial x} + \frac{y}{2}
\frac{\partial}{\partial y},\qquad
G_3  =  \frac{\partial}{\partial t}
\end{gather}
which represent  conformal and self-similar
transformations and time translation. These generators satisfy
the commutation relations
\begin{equation}
[G_3, G_2] = G_3, \qquad
[G_3, G_1] = 2 G_2, \qquad
[G_2, G_1] = G_1.
\end{equation}
By redefining the generators as $G_3=\sigma_1$, $G_2=\sigma_2/2$,
$G_1=\sigma_3$, where
\[
[\sigma_1,\sigma_2] = 2\sigma_1, \qquad
[\sigma_1,\sigma_3] = \sigma_2, \qquad
[\sigma_2,\sigma_3] = 2\sigma_3,
\]
we see that the symmetry algebra is ${\bf sl}(2,\mathbb{R})$.
Actually, it is the characteristic algebra of generalized Ermakov
systems~[5].
We have to note that equation~(22)
can be integrated and the result is
\begin{equation}
\frac{1}{2} (\dot{x} y - x \dot{y})^2 +
\int^{y/x}
\left[u f (u) - u^{-3} g (u)\right] du = \mbox{const}
\end{equation}
which is known as the Ermakov--Lewis invariant
and (23) is also a symmetry
of the Ermakov--Lewis invariant~[4].

If $C$ is a nonzero constant, the function $H(x,y)$
must be the solution of equation (18), that is
\begin{equation}
H =-\frac{h(x/y)}{y^2} + \frac{C}{5} \left(x^2+y^2\right)^{3/2},
\end{equation}
and the function
$\sigma(t)$ must be the solution of
\begin{equation}
\ddot{\sigma}+\frac{4}{5}C\sigma = \zeta,
\end{equation}
where $w = 0$ and $\zeta$ is a constant. This equation can be integrated
easily and the result is
\begin{equation}
\sigma(t)=c_1 e^{\beta t}+c_2 e^{-\beta t} - \frac{\zeta}{\beta^2},
\end{equation}
where $\beta=2i \sqrt{\frac{C}{5}}$ and $c_1$, $c_2$ are constants
of integration. Substituting (29) into (21) we obtain
the symmetries
\begin{gather}
G_1  =  e^{\beta t} \frac{\partial}{\partial t} + \frac{\beta}{2}
e^{\beta t}\left(x\frac{\partial}{\partial x} + y \frac{\partial}{\partial y}\right),
\nonumber\\
G_2  =  e^{-\beta t}\frac{\partial}{\partial t}- \frac{\beta}{2}
e^{-\beta t}\left(x\frac{\partial}{\partial x}+y\frac{\partial}{\partial y}\right),
\qquad
G_3  = - \frac{1}{\beta^2}\frac{\partial}{\partial t}
\end{gather}
with commutation relations
\begin{equation}
[G_1, G_3] =\frac{1}{\beta} G_1, \qquad
[G_1, G_2] =\frac{5}{2} \beta^3 G_3, \qquad
[G_3, G_2] =\frac{1}{\beta} G_2.
\end{equation}
The Lie algebra is again ${\bf sl}(2,\mathbb{R})$
which can be shown easily by choosing $G_1=2\sigma_1/\sqrt{5\beta}$,
$G_2=2\sigma_3/\sqrt{5\beta}$,
$G_3=\sigma_2/2\beta$.
We conclude that Kepler--Ermakov systems have the Lie
algebra ${\bf sl}(2,\mathbb{R})$ if the function
$H$ is in the form of~(27).

By using (27) in (9) we have the equations
\begin{gather}
\ddot{x}  =  -x\left[\frac{C}{5}-\frac{h(x/y)}{y^2\left(x^2+y^2\right)^{3/2}}
\right]+\frac{f(y/x)}{x^3},\nonumber\\
\ddot{y}  =  -y\left[\frac{C}{5}-\frac{h(x/y)}{y^2\left(x^2+y^2\right)^{3/2}}
\right]+\frac{g(y/x)}{y^3}
\end{gather}
possessing the ${\bf sl}(2,\mathbb{R})$ symmetry. These equations can
be considered as the equations of motions of a particle with
unit mass and can be obtained from a Lagrangian function,
\begin{equation}
L=\frac{1}{2}\left(\dot{x}^2+\dot{y}^2\right)-\frac{C}{10}\left(x^2+y^2\right)
  -\frac{1}{2}\left[\frac{f(y/x)}{x^2} + \frac{g(y/x)}{y^2}\right]
  -\Psi(x,y),
\end{equation}
if
\begin{equation}
y^2 f' (y/x) + x^2 g' (y/x)=0,
\end{equation}
where the prime denotes the derivative with respect to the argument and
\begin{equation}
\frac{\partial \Psi}{\partial x}=-\frac{x h(x/y)}{y^2\left(x^2+y^2\right)^{3/2}},\qquad
\frac{\partial \Psi}{\partial y}=-\frac{y h(x/y)}{y^2\left(x^2+y^2\right)^{3/2}},
\end{equation}
so that the force is derivable from the potential function. The function
$\Psi(x,y)$ is integrable if
\begin{equation}
h(x/y)=\frac{C_0 y^2}{x^2+y^2},
\end{equation}
where $C_0$ is an arbitrary constant. As a result we have the
equations of motion
\begin{gather}
\ddot{x}  =  -x\left[\frac{C}{5}-\frac{C_0}{\left(x^2+y^2\right)^{5/2}}\right]
+\frac{f(y/x)}{x^3},\nonumber\\
\ddot{y}  =  -y\left[\frac{C}{5}-\frac{C_0}{\left(x^2+y^2\right)^{5/2}}\right]
+\frac{g(y/x)}{y^3},
\end{gather}
for Kepler--Ermakov systems possessing the ${\bf
sl}(2,\mathbb{R})$ symmetry that are obtained from the Lagrangian
\begin{equation}
L=\frac{1}{2}\left(\dot{x}^2+\dot{y}^2\right)-\frac{C}{10}\left(x^2+y^2\right)
  -\frac{C_0}{3}\left(x^2+y^2\right)^{-3/2}
  -\frac{1}{2}\left[\frac{f(y/x)}{x^2} + \frac{g(y/x)}{y^2}\right]
\end{equation}
with the condition (34).

In polar coordinates the Lagrangian and equations of motion
can be written as
\begin{equation}
L=\frac{1}{2}(\dot{r}^2+r^2\dot{\theta}^2)-\frac{C}{10}r^2
  -\frac{C_0}{3r^3}-\frac{G(\theta)}{2r^2},
\end{equation}
where
$G(\theta)=\sec^2\theta f(\tan\theta)+ \csc^2\theta g(\tan\theta)$,
and
\begin{gather}
\ddot{r}-r\dot{\theta}^2-\frac{G(\theta)}{r^3}+\frac{C}{5}r
-\frac{C_0}{r^4}=0,\\
\frac{d(r^2\dot{\theta})}{dt}+\frac{G'(\theta)}{r^2}=0.
\end{gather}
The Ermakov--Lewis invariant (26) comes from the integration of (41)
and is
\begin{equation}
I=\frac{1}{2}(\dot{\theta}r^2)^2 +
\int^{\tan\theta}
\left[u f (u) - u^{-3} g (u)\right] du.
\end{equation}
Very recently~[7], it was shown that the Ermakov--Lewis invariant
is the result of a Noether dynamical symmetry
for a class of Lagrangian--Ermakov systems.
A Noether symmetry is a Lie point transformation that leaves the
action functional invariant up to an additive constant~[11]. It is
also a Lie symmetry of the corresponding Euler--Lagrange equations.
If
\begin{equation}
{\bf X} = \xi \frac{\partial}{\partial t} + \eta^{a}
\frac{\partial}{\partial q^{a}} + (\dot{\eta}^{a}-\dot{q}^{a}\dot{\xi})
\frac{\partial}{\partial \dot{q}^{a}},
\end{equation}
is the generator of a Noether symmetry, then
\begin{equation}
\varphi=\xi\left[\dot{q}^{k}\frac{\partial L}{\partial \dot{q}^{k}}-L\right]
        -\eta^{k}\frac{\partial L}{\partial \dot{q}^{k}}
        +\Lambda(q^i,t)
\end{equation}
is a first integral that satisfies ${\bf X}\varphi=0$.
For the Lagrangian in (38) a Noether symmetry generator is
$G_3$ and the first integral determined by (44) is the Hamiltonian.

Finally we calculate the generator of Ermakov invariant for the
Lagrangian (38). In general, if the Lagrangian $L$ and a first
integral $\varphi$ of a dynamical system are known one can determine
the corresponding symmetry of dynamical character~[11]. If
\begin{equation}
{\bf X} = \xi(q^{k},\dot{q}^{k},t)\frac{\partial}{\partial t} +
\eta^{a}(q^{k},\dot{q}^{k},t)
\frac{\partial}{\partial q^{a}} + \dot{\eta}^{a}(q^{k},\dot{q}^{k},t)
\frac{\partial}{\partial \dot{q}^{a}},
\end{equation}
is the generator of a dynamical symmetry, then there exist a first
integral $\varphi$ such that
\begin{equation}
\frac{\partial^{2} L}{\partial \dot{q}^{a}\partial \dot{q}^{b}}
(\eta^{a}-\dot{q}^{a}\xi)=-\frac{\partial\varphi}{\partial \dot{q}^{b}}
\end{equation}
holds. Conversely, if $\varphi$ is a first integral, then
$\eta^{a}-\dot{q}^{a}\xi$ determines a dynamical symmetry which is
called a Cartan symmetry~[11]. We can now use (46) to determine the
dynamical symmetry corresponding to the Ermakov invariant (26) and
then have
\begin{gather}
{\bf X} =(x\dot{y}-y\dot{x})\left(y\frac{\partial}{\partial x}-
          x\frac{\partial}{\partial y}\right)
 + \left[\dot{y}(x\dot{y}-y\dot{x})+\frac{x}{y^2}g(y/x)-\frac{y^2}
          {x^3}f(y/x)\right]\frac{\partial}{\partial \dot{x}} \nonumber\\
\phantom{{\bf X} =}{}- \left[\dot{x}(x\dot{y}-y\dot{x})+\frac{x^2}{y^3}g(y/x)-\frac{y}
          {x^2}f(y/x)\right]\frac{\partial}{\partial \dot{y}},
\end{gather}
which is not a point symmetry [11] because it is not possible to add it
a multiple of ${\bf A}$ and thereby to get rid of all derivatives
in the coefficients of $\partial_x$ and
$\partial_y$.

\section{Conclusion}
In this work we concentrated on the Kepler--Ermakov systems, with
$w(t)=0$, possessing the ${\bf sl}(2,\mathbb{R})$ symmetry.
We found that these systems are usual
Ermakov systems with the frequency function depending on the dynamical
variables. We obtained the Lagrangian for these systems and calculated
the Cartan symmetry corresponding to the Ermakov invariant.

\subsection*{Acknowledgements}
We would like to thank S~Yu~Sakovich for
his careful reading and suggestions on the manuscript. We also
gratefully acknowledge the editor, P~G~L Leach, and the referee
for valuable comments and suggestions.
This work is supported in part by the
Scientific and Technical Research Council of Turkey (TUBITAK).

\label{karasu-lastpage}

\end{document}